\documentstyle[12pt,epsf]{article}

\newcommand{\mrm}[1]{\mbox{\rm #1}}

\newcommand{\beq}{\begin{equation}}
\newcommand{\eeq}{\end{equation}}
\newcommand{\nn}{\nonumber}
\newcommand{\bea}{\begin{eqnarray}}
\newcommand{\eea}{\end{eqnarray}}

\newcommand{\Eq}[1]{Eq.~(\ref{#1})}
\newcommand{\gsim}{\ \rlap{\raise 2pt\hbox{$>$}}{\lower 2pt \hbox{$\sim$}}\ }
\newcommand{\lsim}{\ \rlap{\raise 2pt\hbox{$<$}}{\lower 2pt \hbox{$\sim$}}\ }

\newcommand{\np}[1]{Nucl. Phys. {\bf #1}}
\newcommand{\pl}[1]{Phys. Lett. {\bf #1}}
\newcommand{\pr}[1]{Phys. Rev. {\bf #1}}
\newcommand{\prl}[1]{Phys. Rev. Lett. {\bf #1}}
\newcommand{\zp}[1]{Z. Phys. {\bf #1}}

\newcommand{\ptp}[1]{Prog. Theor. Phys. {\bf #1}} 
\newcommand{\arns}[1]{Ann. Rev. Nucl. Sci. {\bf #1}}

\makeatletter
\if@twoside
\else \oddsidemargin 00pt \evensidemargin 00pt
\fi
\let\@eqnsel = \hfil

\expandafter\ifx\csname mathrm\endcsname\relax\def\mathrm#1{{\rm #1}}\fi
\@ifundefined{mathrm}{\def\mathrm#1{{\rm #1}}}{\relax}

\makeatother

\begin{document}

\thispagestyle{empty}
\null
\hfill FTUV/96-70, IFIC/96-79

\hfill hep-ph/9702337

\vskip 1.5cm

\begin{center}
{\Large \bf      
SPONTANEOUS CP VIOLATION  \\ AND THE $B^0$ SYSTEM 
\par} \vskip 2.em
{\large		
{\sc G. Barenboim$^{1,2}$, J. Bernab\'eu$^1$ and M. Raidal$^{1,2}$
}  \\[1ex] 
{\it $^1$ Departament de F\'\i sica Te\`orica, Universitat 
de Val\`encia}\\ 
{\it $^2$ IFIC, Centre Mixte Universitat 
de Val\`encia - CSIC} \\
{\it E-46100 Burjassot, Valencia, Spain} \\[1ex]
\vskip 0.5em
\par} 
\end{center} \par
\vfil
{\bf Abstract} \par
We investigate effects of spontaneous breakdown of CP in 
$B^0_{d,s}-\overline{B^0}_{d,s}$ systems in left-right symmetric models.
Assuming that the left-right contribution to the
$B^0-\overline{B^0}$ matrix element $M_{12}$  can be at most 
equal to the standard model one
we obtain a new lower bound, $M_H\gsim 12$ TeV, on the
flavour changing Higgs boson mass. Most importantly, the convention
independent parameter $\mrm{Re}(\overline{\epsilon}_B),$ which measures
the amount of $\Delta B=2$ CP violation, can be enhanced by a factor of four
or more for $B^0_d$ and almost by two orders of magnitude for $B^0_s$ systems
when compared with the Standard Model predictions. 
Therefore, interesting possibilities
to observe indirect CP violation in the $B$ system 
are open in the planned facilities.
   
\par
\vskip 0.5cm
\noindent December 1996 \par
\null
\setcounter{page}{0}
\clearpage

\section{Introduction}

The success of the Standard Model (SM) teaches us very little about
what to expect at energies higher than the left-handed gauge boson
mass. An interesting, viable extension of the SM is the
left-right symmetric $SU(2)_R \times SU(2)_L \times U(1)_{B-L}$ 
model \cite{uno}. 
Besides giving rise to a new energy scale 
the most interesting character of the model is the spontaneous
breaking of parity invariance. In such a model of spontaneous 
parity violation, it seems natural to consider the possibility 
that CP is also broken spontaneously \cite{dos,tres,chang,eg1,frere,seis}.
More interestingly, it has been shown \cite{tres} 
that, to implement this
symmetry breaking, it is not necessary to extend the Higgs boson
sector while keeping flavour changing neutral currents under control.

Although the SM has passed all the tests in the kaon system 
\cite{cuatro,ocho},
intriguing hints of other plausible explanations emerge from
cosmological considerations of the baryon to photon ratio in the
Universe \cite{cinco}. 
The left-right symmetric model can equally well explain
the existing measurements of CP violation in the neutral kaon system
\cite{seis}. However, by studying  kaons alone it is very difficult
to distinguish between different mechanisms of CP violation. 
Complementary tests of the origin of CP violation 
will be provided by the studies of $B$- systems in 
planned facilities \cite{bfac} which will start to operate in the 
near future.   Unfortunately, the convention independent 
parameter $\overline{\epsilon},$
which measures the amount of $\Delta B=2$ CP violation, is predicted
to be very small in the SM setting extremely strong requirements 
on the performance of these machines. In fact, it is  possible that 
even the SM value of $\mbox{Re}(\overline{\epsilon})$ for $B^0_d,$ 
which is expected to be an order of
magnitude larger than the one for $B^0_s,$ will not be 
achievable in these experiments.       
Therefore, measurements of large $\overline{\epsilon}$ would inevitably
mean the discovery of physics beyond the SM.
It is for this reason why the exploration of CP
violation in the $B$ system is so crucial.

The spontaneous violation of CP in $B^0_{d,s}$
systems have been studied  earlier  \cite{siete}. 
In the present work we shall re-consider mixings and CP violation in the
 B systems 
in view of the progress in our understanding
of the hadronic matrix elements and implementation of the spontaneous 
breakdown of CP in left-right symmetric models, not to
mention the improvements in the precision of experimetal data such
as the strong coupling constant  or top quark mass.  
As will be seen,  our results indicate that an enhancement of about
an order of magnitude compared with the SM prediction is possible leading 
to observable effects in the future experiments.

The outline of the paper is the following. In Section 2 we discuss
the
spontaneous breakdown of CP in left-right symmetric models. In Section 3
we present the general formalism for the mixing and CP violating
parameters in $B^0-\overline{B}^0$ system and in Section 4 we derive the 
parameters in our model. Numerical 
results are presented in the same Section.
Our conclusions are drawn in Section 5.

\section{Spontaneous CP violation in the left-right symmetric model}

Here we present a brief review of the minimal $SU(2)_R \times SU(2)_L 
\times U(1)_{B-L}$ model with a discrete left-right symmetry. 
The purpose is to set the stage for the model we consider and 
establish some notation. A more detailed description of the model 
exists in the literature \cite{uno}.
The fermionic sector of the model contains three generations
of quarks which we denote as $\Psi_{iL} \equiv (u_i , d_i)_L$
in the representation ($\frac{1}{2}$, 0, $\frac{1}{3}$) and
$\Psi_{iR} \equiv (u_i , d_i)_R$ in the representation 
(0, $\frac{1}{2}$, $\frac{1}{3}$), 
where i=1, 2, 3 denotes the corresponding generation. 
The leptonic sector does not concern us here. The Higgs boson sector
contains a bidoublet $\Phi$ in the representation 
($\frac{1}{2}$, $\frac{1}{2}$, 0) and two triplets, 
 $\Delta_L$ and $\Delta_R,$ in the representations 
(1, 0, 2) and  (0, 1, 2), respectively.
They can be written as
\bea
\Phi \equiv  \frac{1}{2} \pmatrix { \phi_1^0 & \phi_1^+ \cr
	\phi_2^- & \phi_2^0 \cr} \; \; \; , \; \;\;
\Delta_L \equiv \pmatrix { \delta_L^+/\sqrt{2} & \delta^{++}_L \cr
\delta_L^0 & -\delta_L^+/\sqrt{2} \cr}.
\eea
In order to have parity as a spontaneously broken symmetry,
a discrete left-right symmetry is imposed:
\bea
\Psi_{iL} \longleftrightarrow \Psi_{iR} \; \; \; ,
\; \;\; \Delta_L  \longleftrightarrow \Delta_R \;\;\;
, \;\;\; \Phi  \longleftrightarrow \Phi^\dagger.
\label{disc}
\eea
The most  general Yukawa interaction invariant under (\ref{disc})
can be written as
\bea
\cal{L}_Y &=& \sum_{i,j=1}^3 \left(
f_{ij} \bar{\Psi^i}_L \Phi  \Psi^j_R  \, +  \, 
g_{ij} \bar{\Psi^i}_L  \tilde{\Phi}  \Psi^j_R
\right) \; + \; h.c.,
\label{fandg}
\eea
where $\tilde{\Phi} \equiv \tau_2 \Phi^* \tau_2 .$
After  symmetry breaking,  the vacuum expectation value (vev) of
$\Phi$ can be written as
\bea
\langle \Phi \rangle  = 
\pmatrix {k_1 e^{i\alpha^\prime} & 0 \cr 
0 & k_2  e^{i \alpha } \cr} \; .
\label{higgs}
\eea
The quark mass matrices generated by Eqs. (\ref{fandg}), (\ref{higgs}) are
\bea
\hat{M}^u = \hat{f} k_1  e^{i \alpha^\prime} + \hat{g} k_2 e^{-i \alpha} 
\; ,\nn \\
\hat{M}^d = \hat{g} k_1  e^{-i \alpha^\prime} + \hat{f} k_2 e^{i \alpha} 
\; ,
\label{qmass}
\eea
where $\hat{M}^u$ ($\hat{M}^d$) is the up (down) quark mass
matrix and the hats denote $3\times3$ matrices.
As a result of the left-right discrete symmetry, $\hat{f}$ and
$\hat{g}$ must be Hermitian. However, $\hat{M}^u$ and $\hat{M}^d$
are not Hermitian. In order to obtain the left-
and right-handed Cabbibo-Kobayashi-Maskawa 
(CKM) matrices $V_L$ and $V_R,$ respectively, 
they must be diagonalized by the usual bi-unitary transformation. 
In general, there is no simple relationship between 
$V_L$ and $V_R$. 

The vevs of the Higgs bosons,
$\langle \Phi \rangle$ and $\langle \Delta_{L,R} \rangle\equiv v_{L,R}$, 
generate the following mass matrix for the charged gauge bosons 
\bea
\pmatrix { \frac{1}{2} g^2 (k_1^2 + k_2^2) + g^2 v_L ^2 & 
-g^2 k_1 k_2 e^{-i (\alpha - \alpha^\prime)} \cr
-g^2 k_1 k_2 e^{i (\alpha - \alpha^\prime)} &
 \frac{1}{2} g^2 (k_1^2 + k_2^2) + g^2 v_R ^2 \cr}.
\label{mix}
\eea
Experimentally we know that $v_R^2 \gg k_1^2 , k_2^2 \gg v_L^2, $ 
which imply that the $W_1^\pm$ mass is given to a good approximation
by $M_1 \simeq \frac{1}{2} g^2 (k_1^2 +k_2^2) $, and similarly
$M_2 \simeq \frac{1}{2} g^2 v_R^2$. The $W_1^\pm-W_2^\pm$ 
mixing angle $\xi$ is small, 
\bea
\xi \approx \frac{2 k_1 k_2}{k_1^2 + k_2^2} \frac{M_1^2
}{M_2^2} \; ,\nn 
\eea
as required by the low energy phenomenology \cite{lang,meie2}.
The most stringent lower bound on the new gauge boson mass
$W_2\gsim 1.6$ TeV  derives from the analysis of the $K_L-K_S$ 
mass difference \cite{beall}. 
However, this bound depends quite strongly 
on low energy QCD and 
different assumptions used in the literature
\cite{meie2}.  Within the present errors, it can be as low as $M_2\sim 900$ 
GeV, not too far from the Tevatron bound $M_2\gsim 652$ GeV \cite{tevatron}. 
There are two neutral flavour changing Higgs bosons in the model.
To suppress their interactions, the lower bound $M_H\gsim 10$ TeV
has been derived \cite{eg1}.

The model we want to analyse is the one we have described, except
that we impose CP as a spontaneously broken symmetry. 
In this model it is not necessary to introduce any extra Higgs 
multiplet in order to break CP spontaneously  \cite{tres}.
In this sense, it is more natural to discuss spontaneous CP violation
in models with the gauge group $SU(2)_R \times SU(2)_L \times 
U(1)_{B-L}$ rather than in the standard $SU(2)_L \times U(1)$ models
in which extra Higgs multiplets are needed  to generate the
spontaneous CP violation. 
A direct consequence of imposing CP as a spontaneously
broken symmetry, together with
the discrete left-right symmetry (\ref{disc}), is that the Yukawa coupling
matrices $\hat{f}$ and $\hat{g}$ in Eq. (\ref{fandg}) must
be real symmetric. The only complex parameter in the mass matrices
$\hat{M}^{u,d}$ is a complex phase in
$\langle \Phi \rangle$ that we will discuss now.

In order to break CP spontaneously, we have to look for a complex 
vev of the Higgs boson.  
The vev $\langle \Phi \rangle$ of Eq. (\ref{higgs}) breaks the
$U(1)_{L-R}$ symmetry with the generator $I_{3L} - I_{3R}$. We can use
this $U(1)$ symmetry to shift the phases between the $k_1$ and $k_2$ 
component of the $\langle \Phi \rangle$. As a result we can choose
$\alpha^\prime$ to be zero without loss of generality.
Similarly, also $v_R $ can be made real
by using the $U(1)_{B-L}$ symmetry that it breaks. Working 
in the limit of $k_2 \sim v_L
\sim 0$, as is usually assumed in the literature, 
we would find that there is no CP violation at all.
A small, complex nonzero $v_L $ will generate CP violation in the
leptonic sector, but not in the pure hadronic sector.

\section{The  $B^0$-$\overline{B}^0$ system}

The flavour quantum numbers are not conserved  by weak interactions. Thus
a $B^0$ state can be transformed into its antiparticle $\overline{B}^0$.
As a consequence, the flavour eigenstates $B^0$ and $\overline{B}^0$ are
not mass eigenstates and do not follow an exponential decay law.

Let us consider an arbitrary mixture of two flavour states
\bea
\mid \psi(t) \rangle \; = \; a(t) \, \mid B^0 \rangle \; + \;
b(t)  \, \mid \overline{B^0} \rangle \; \equiv \;
\pmatrix {a(t) \cr b(t) \cr}.
\eea
Its time evolution is governed by the equation
\bea
i \, \frac{d}{dt}  \mid \psi(t) \rangle \; = \; \cal{M}\, \mid
\psi(t) \rangle,
\eea
where $\cal{M}$ is called the $B^0 - \overline{B^0}$ mixing matrix.
Assuming CPT symmetry to hold, this can be written as
\bea
\cal{M} = \pmatrix { M - \frac{i}{2} \Gamma & M_{12} - \frac{i}{2} 
\Gamma_{12} \cr M_{12}^* - \frac{i}{2} \Gamma_{12}^* &
M- \frac{i}{2} \Gamma \cr}\; .
\eea
The diagonal elements $M$ and $\Gamma$ are real parameters, which
would correspond to the mass and width of the neutral mesons in the
absence of mixing. The off-diagonal entries contain the dispersive
and absorptive parts. If CP were an exact symmetry, $M_{12}$ and
$\Gamma_{12}$ would also be real.
The physical eigenstates of $\cal{M}$ are,
\bea
\mid B_\pm \rangle \; = \; \frac{1}{\sqrt{\mid p\mid^2 +
\mid q \mid^2}} \left( p \mid B^0 \rangle \, \pm \, 
q \mid \overline{B^0} \rangle \right) \; ,
\eea
with
\bea
\frac{q}{p}  = \frac{1 - \overline{\epsilon}}{1 + \overline{\epsilon}} =
\left( \frac{M_{12}^* - \frac{i}{2} \Gamma_{12}^*}{
M_{12} - \frac{i}{2} \Gamma_{12}} \right)^{\frac{1}{2}} \; .
\eea
If $M_{12}$ and $\Gamma_{12}$ were real, then $\frac{q}{p} =1$
and $\mid B_\pm \rangle$ would correspond to the CP even and
CP odd eigenstates.

Note that if the $B^0 - \overline{B}^0$ violates CP, the two mass
eigenstes are no longer orthogonal, and we can define
\bea
\langle B_- \mid B_+ \rangle = \frac{ \mid p \mid^2 - \mid q \mid^2}
{\mid p \mid^2 +\mid q \mid^2} = \frac{2 \mbox{Re}(\overline{\epsilon})}
{1 +\mid \overline{\epsilon} \mid^2} \approx 2 
\mbox{Re}(\overline{\epsilon}) \; ,
\label{cpp}
\eea
which is a convention independent measure of CP violation.
Notice that we have used the shorthand notation $\overline{\epsilon} 
\equiv \overline{\epsilon}_B$.

The time evolution of a state which was originally produced
as a $B^0$ or a $\overline{B}^0$ is given by
\bea
\pmatrix { \mid B^0(t) \rangle \cr \mid \overline{B}^0(t) \rangle \cr}
= \pmatrix { g_1(t) & \frac{q}{p} g_2(t) \cr \frac{p}{q} g_2(t) &
g_1(t) \cr} \pmatrix { \mid B^0 \rangle \cr 
\mid \overline{B}^0 \rangle \cr} \; ,
\eea
where
\bea
\pmatrix { g_1(t) \cr g_2(t) \cr} = e^{i M t} e^{- \Gamma t / 2}
\pmatrix { \cos( \Delta M -\frac{i}{2} \Delta\Gamma)\frac{t}{2} \cr
-i \sin( \Delta M -\frac{i}{2} \Delta\Gamma)\frac{t}{2} \cr} \; ,
\eea
with
\bea
\Delta M = M_{B_+} - M_{B_-} \;\; \; \mbox{and} \;\;\;
\Delta\Gamma = \Gamma_{B_+} - \Gamma_{B_-}\; .
\eea
The mass difference $\Delta M$, and the difference of the decay widths
can be calculated from the general expressions \cite{bursl}
\bea
\Delta M &=&  \sqrt{2} \left( \left(d_1^2 + d_2^2 \right)^{\frac{1}{2}} 
+ d_1 \right)^{\frac{1}{2}} \; ,\nn \\
\Delta\Gamma &=& -2 \sqrt{2}\, \mbox{sgn}(d_2) 
\left( \left(d_1^2 + d_2^2 \right)^{\frac{1}{2}} 
- d_1 \right)^{\frac{1}{2}} \; ,
\eea
with
\bea
d_1 &=& \mid M_{12} \mid^2 - \frac{1}{4} \mid \Gamma_{12} \mid^2 
\; ,\nn \\
d_2 &=& \mbox{Re}( M_{12} \Gamma_{12}^* ) \; .
\eea
The main difference between the $K^0 - \overline{K}^0$ and
$B^0 - \overline{B}^0$ system stems from the different kinematics
involved. The light kaon mass only allows the hadronic decay modes
$K^0 \longrightarrow 2 \pi$ and $K^0 \longrightarrow 3 \pi$. Since
CP$\mid \pi\pi \rangle = + \mid \pi \pi \rangle$, the CP even kaon 
state decays into 2$\pi$ whereas the CP odd one decays into the 
phase space suppressed 3$\pi$ mode. Therefore, there is a large
lifetime difference and we have a short-lived $\mid K_S \rangle
\equiv \mid K_- \rangle \approx \mid K_1 \rangle + \overline{\epsilon}_K
\mid K_2 \rangle $ and a long lived  $\mid K_L \rangle
\equiv \mid K_+ \rangle \approx \mid K_2 \rangle + \overline{\epsilon}_K
\mid K_1 \rangle $ kaon, with $\Gamma_{K_L} \ll \Gamma_{K_S}$.
One finds experimentally that
\bea
\Delta \Gamma_{K^0} \approx - \Gamma_{K_S} \approx - 2 \Delta M_{K^0}
\eea
In the $B$ system, there are many open decay channels and a large
part of them are common to both mass eigenstates. Therefore, the
$\mid B_\pm \rangle$ states have a similar lifetime, i.e., 
$\Delta \Gamma_{B^0} \ll \Gamma_{B^0}$. Moreover,
whereas the $B^0 - \overline{B}^0$ mixing is dominated by the top in the
box diagram, the decay amplitude gets obviously its main contribution
from the $b \rightarrow c$ transition. Thus,
\bea
\frac{\Delta\Gamma_B}{\Delta M_B} \sim \frac{m_b^2}{m_t^2} \ll 1.
\eea

\section{Left-right contributions to the $B$ system}

Let us examine now the left-right contribution to the $B$ system. 
The charged current Lagrangian is given by 
\bea
\cal{L}_{cc} &=& \frac{g}{\sqrt{2}} \; \overline{u} \left( \cos\xi V_L 
\gamma^\mu P_L\, - \, e^{i \alpha} \sin\xi V_R \gamma^\mu P_R \right) d \;
W_{1\mu} \; \; +  \nn \\
&&\frac{g}{\sqrt{2}} \; \overline{u} \left( 
e^{-i\alpha} \sin\xi V_L 
\gamma^\mu P_L\, +\, \cos\xi V_R \gamma^\mu P_R \right) d \;
W_{2\mu} \;\; + \;\; \mbox{h.c.},
\eea
with
\bea
P_{L,R} \equiv \frac{(1 \mp \gamma_5)}{2} \; ,
\nn \eea
where $W_1$ ($W_2$) is the charged vector boson field with the mass
$M_1$ ($M_2$).

The  left-right model contributions to $\langle B^0 \mid H \mid 
\overline{B^0} \rangle$ are shown in Fig. 1.  
As in the kaon
system, multiplicatively renormalizable operators of the type
\bea
O_{LL} &=& \overline{b}\gamma^{\mu} P_L l \overline{b}\gamma_{\mu}
P_L l  \; ,\nn \\
O_S &=& \overline{b} P_L l \overline{b}P_R l  \; ,\nn \\
O_{\tilde{V}} &=& \overline{b} \gamma^\mu P_L l \overline{b} 
\gamma_\mu P_R l + \frac{2}{3} O_S,
\eea
with $l$=d- or s-quark, appear in the effective Hamiltonian. Their matrix
elements, evaluated at an energy scale $m_b(m_b) = 4.3$ GeV,
are 
\bea
\langle B^0 \mid O_{LL} \mid \overline{B^0} \rangle & = &
\frac{8}{3} B_B f_B^2 m_B^2 \; , \nn \\
\langle B^0 \mid O_S \mid \overline{B^0} \rangle & = &
\left( \frac{m_B^2}{(m_b + m_l)^2} + \frac{1}{6} \right) 
B'_B f_B^2 m_B^2 
\; ,\nn \\
\langle B^0 \mid O_{\tilde{V}} \mid \overline{B^0} \rangle & = &-
\frac{8}{9} B''_B f_B^2 m_B^2 \; ,
\eea
where $m_B=5.28$ GeV is the $B^0$ mass, $f_B$ denote the $B$ meson
decay  constant and the
factors $B_B,$ $B'_B,$ and $B''_B$ take into account 
deviations of the actual values of
the hadronic matrix elements  from the vacuum insertion approximation.
At present, only the combination $\sqrt{B_B}f_B=200\pm50$ MeV 
is known \cite{ocho}.
In the following we assume that $B_B=B'_B=B''_B$
which should be a good approximation for our numerical estimates.

With these hadronic matrix elements 
the SM contribution (Fig. 1a)  is given by \cite{ocho} 
\bea
M_{12}^{LL} \simeq \frac{G_F^2}{12 \pi^2} B_B f_B^2 m_B M_1^2
\eta_2^{(B)} (\lambda_t^{LL})^2  S(x_t)\; ,
\eea
where  $\eta_2^{(B)} \simeq 0.55 $ is a QCD correction factor,
\bea
\lambda_t^{AB} = V_{A,tl} V_{B,tb}^* \; \; \; \; \;(A,B=L,R),
\eea 
and \cite{lim}
\bea
S(x) =  -\frac{3}{2} \left(\frac{x}{1-x} \right)^3 \ln(x) + 
x \left( \frac{1}{4} + \frac{9}{4} \frac{1}{(1-x)}  - \frac{3}{2}
\frac{1}{(1-x)^2} \right).
\eea
The tt exchange dominates not only the SM contribution but also
the left-right contribution. For the transverse  $W_1$ and $  W_2$ 
one gets \cite{siete}  
\bea
M_{12}^{W_1 W_2} \simeq \frac{G_F^2}{\pi^2} M_1^2 \beta 
B_B f_B^2 m_B \left( \left(\frac{m_B}{m_b + m_l} \right)^2 + \frac{1}{6}
\right) \lambda_t^{RL} \lambda_t^{LR} \eta_1^{LR}
F_1(x_t, x_b, \beta),
\eea 
and for 
the $S_1 W_2$ contribution in Fig. 1c ($S_i$ denoting the Goldstone boson
which becomes the longitudinal component of $W_i$)
\bea
M_{12}^{S_1 W_2} \simeq - \frac{G_F^2}{4 \pi^2} M_1^2 \beta 
B_B f_B^2 m_B \left( \left(\frac{m_B}{m_b + m_l} \right)^2 + \frac{1}{6}
\right) \lambda_t^{RL} \lambda_t^{LR} \eta_2^{LR}
F_2(x_t, x_b, \beta).
\eea 
Here  $\beta = M_1^2 /M_2^2$, $x_i = m_i^2 /M_1^2$ and  
$\eta_1^{LR}$ and $ \eta_2^{LR} $ are the QCD coefficients.
Following Ref. \cite{eg1} and assuming $\alpha_s(M_Z)=0.118$
 and quark masses as given in Ref. \cite{pdb} 
we obtain 
$\eta_1^{LR}=1.83$ and $ \eta_2^{LR}=1.66.$
The function $F_1$ is given by
\bea
F_1(x_t, x_b, \beta) = \frac{ x_t}{(1-x_t \beta)
(1-x_t)} \int_0^1 d\alpha \, {\sum_k}^\prime \ln \mid
\Lambda_k(\alpha,\beta)\mid,
\eea
where
\bea
{\sum_k}^\prime  = \sum_{1,2} - \sum_{3,4},
\eea
and the functions $\Lambda_k(\alpha, \beta)$ are defined by
\bea
\Lambda_1(\alpha, \beta) &=&  x_t  - x_b \alpha (1 - \alpha) \; , \nn \\
\Lambda_2(\alpha, \beta) &=& 1 - \alpha + \frac{\alpha}{\beta} -
x_b \alpha (1 - \alpha) \; , \nn \\
\Lambda_3(\alpha, \beta) &=& x_t (1 - \alpha) + \frac{\alpha}{\beta}
- x_b \alpha (1 -\alpha) \; ,\nn \\
\Lambda_4(\alpha, \beta) &=& 1 - \alpha + x_t \alpha
- x_b \alpha (1 -\alpha) \; .
\eea
Analogously,
\bea
F_2(x_t, x_b, \beta) = \frac{ 2 x_t }{(1-x_t \beta)
(1-x_t)} \int_0^1 d\alpha \, {\sum_k}^\prime  \Lambda_k(\alpha,\beta) 
\ln \mid\Lambda_k(\alpha,\beta) \mid \; .
\eea
Among the three diagrams containing the unphysical scalars $S_i$, the ones
which contain the $W_1 S_2$ (Fig. 1d) and $S_1 S_2$  (Fig. 1e) boxes are
of order $\beta^2$ and can be neglected even for the intermediate top
quark. Also the box diagram which
contains two $W_2$ (Fig. 1f) can be neglected due to the same reason.

The tree level contributions are mediated by the flavour changing
neutral Higgs bosons $\phi_1$ and $\phi_2$ and their contribution
is \cite{egn}
\bea
M_{12}^H \simeq - \frac{\sqrt{2} G_F}{M_H^2} m_t^2(m_b) B_B  f_B^2 
m_B  \left( \left(\frac{m_B}{m_b + m_l} \right)^2 + \frac{1}{6}
\right) \lambda_t^{RL} \lambda_t^{LR} \; ,
\eea
where we have assumed a common Higgs mass $M_H$. 
Note that these flavour changing neutral contributions are suppressed
by the same factors $\lambda_t$ as the gauge mediated amplitudes. 
One loop Higgs
contributions to $M_{12}$ including also charged Higgs bosons
are suppressed as far as $M_{H} \gsim M_2$. 
Considering the kaon system, 
this is a quite natural assumption.  In order to make
the left-right contribution to the kaon mass difference smaller than the 
experimental value, the bounds $M_2 \gsim 1.6$ TeV and 
$M_H \gsim 10$ TeV have emerged.

Now we can demonstrate the
relevance of the left-right contribution to the mixing of 
$B^0-\overline{B}^0.$ Using the formulae above we have performed 
a numerical computation resulting 
\bea
\frac{\mid M_{12}^{W_1 W_2} + M_{12}^{S_1 W_2} + M_{12}^{H} \mid}
{\mid M_{12}^{LL} \mid} = F(M_2) \left(\frac{1.6 \mbox{TeV}}
{M_2} \right)^2 +  \left(\frac{12 \mbox{TeV}}
{M_H} \right)^2 \; ,
\label{m12}
\eea
where we have assumed $\mid\lambda_t^{RL}\mid \,= \,
\mid\lambda_t^{LR}\mid \,= \, \mid\lambda_t^{LL}\mid, $
which is guaranteed in the left-right  model with the
manifest left-right symmetry (\ref{disc}). 
The function $F(M_2)$ presents the non-trivial
dependence of the left-right gauge boson box contribution $M_{12}^{LR}$ 
on $M_2$, and is plotted in Fig. 2.
Comparing the numerical values one can see that for Higgs masses close 
to its lower bound the Higgs exchange dominates over the box contribution.  
Since $F(M_2)$ is an increasing function of 
$M_2$ it follows from Eq. (\ref{m12}) that scaling  
the masses $M_2$ and $M_H$ up by the same factor the gauge boson contribution 
becomes relatively more important.
It is important to notice that the 
left-right box and Higgs contributions add up constructively.
With the $W_R$ and Higgs masses close to their present lower bounds 
the modulus  of the left-right contribution to $M_{12}$  
is comparable in magnitude with  the SM one. 
If we assume as in Ref. \cite{eg1} 
that  $\mid M_{12}^{LR}\mid \lsim \mid M_{12}^{LL}\mid $
is allowed experimentally  
we can obtain from Eq. (\ref{m12}) a new lower bound, 
\bea
M_H\gsim 12\; \mbox{{\rm TeV}} ,
\eea
on the Higgs boson mass.   

However, due to the unknown relative phase between the SM and 
left-right contributions, substantial cancellation between them  
is possible. There are all together six phases present in 
the left- and right-handed CKM matrices which may cause destructive 
interference. 
If the CMK phases are small, as  is expected in our model (we will
see in the next section  
that they are proportional to $r \sin\alpha$ which is bounded to 
be a small quantity),
 this cancellation may appear naturally. 
In conclusion, by studying  the $B$ meson mixing alone  
it is difficult, if not impossible, 
to distinguish the left-right model from the  
SM. We move now to investigate also the CP violating observables. 

The result above does not imply  that in the left-right symmetric 
model there is the same enhancement factor  
also for the CP violating observables. 
For the purpose of studying CP violation 
we need to discuss $\Gamma_{12}$. Clearly, tree level Higgs 
exchange does not contribute and all terms coming from
the left-right symmetric model have a suppresing $\beta$ factor 
when compared to the SM contribution. Therefore, the left-right
contribution is completely negligible and to a good approximation
$\Gamma_{12} \approx \Gamma_{12}^{LL}$.
To calculate the transition width we use the quark box diagram 
approximation which is expected to give a good estimate for the $B$ 
system.\footnote{We thank A. Pich for clarifying discussions on  this point.} 
In this approximation $\Gamma_{12}$
can be written in the form \cite{lus}
\bea
\Gamma_{12} = \frac{G_F^2 f_B^2 B m_B^3}{8 \pi} \left(
(\lambda^{LL}_u)^2 X_{uu} + (\lambda^{LL}_c)^2 X_{cc} + 
2 \lambda_c^{LL} \lambda_u^{LL} X_{uc}
\right),
\eea
where the functions $X$ will be defined later. 
In the three generation standard model, as well as in the minimal
left-right  symmetric model, the unitarity of the CKM matrix $V_L$
entails
\bea
\lambda_u^{LL} + \lambda_c^{LL} + \lambda_t^{LL} = 0,
\eea 
and, consequently, 
\bea
\Gamma_{12} = \frac{G_F^2 f_B^2 B m_B^3}{8 \pi} \left[
(\lambda_u^{LL})^2 \left( X_{uu} - X_{uc} \right) + (\lambda_c^{LL})^2 
\left( X_{cc} - X_{uc} \right) + (\lambda_t^{LL})^2  X_{uc}
\right].
\label{g12}
\eea
Neglecting the mass of the up quark, one has $\left( x = m_c^2/m_B^2 
\; , \; y = m_b^2/m_B^2 \right)$ \cite{hh}
\bea
X_{uu} & = &\eta^{(s)} y, \nn \\
X_{cc} &=& \eta^{(s)} y \sqrt{ 1 - \frac{4x}{y}} - \eta^{(a)} \frac{2}{N}
	x \sqrt{1 - 4x}, \nn \\
X_{uc} &=&  \eta^{(s)} y \left( 1+ \frac{x}{y} \right)
	 \left( 1- \frac{x}{y} \right)^2 - \eta^{(a)} \frac{1}{N} x 
\left(1-x \right)^2 ,
\eea
where
\bea
\eta^{(s)} & = &\frac{1}{2} \left(1+ \frac{1}{N}\right) c_+^2 +
	 \frac{1}{2} \left(1- \frac{1}{N}\right) c_-^2, \nn \\
\eta^{(a)} &=& \left( \frac{N+1}{2}\right)^2 c_+^2 + 
	\left( \frac{N-1}{2}\right)^2 c_-^2 - \frac{N^2-1}{2}
c_+ c_- ,
\eea
with \cite{glee}
\bea
c_+ = \left( c_- \right)^{-\frac{1}{2}} = \left( \frac{\alpha_s(m_b)}
{\alpha_s(m_W)}\right)^{-\frac{6}{23}}.
\eea
Here $N=3$ denotes the number of colours.
Numerically the third term in Eq. (\ref{g12}) 
dominates over the first two  and we obtain
\bea
\Gamma_{12}  \approx \frac{G_F^2 f_B^2 B m_B^3}{8 \pi} 
(\lambda_t^{LL})^2  X_{uc}. 
\eea
In agreement with the discussion in Section 3, the ratio
\bea
\left| \frac{\Gamma_{12}}{M^{LL}_{12}} \right| = 
\frac{3 m_B^2 \pi}{2 \eta^{(B)}_2 M_1^2} \left| \frac{X_{uc}}{S(x_t)} \right|
\eea
is, indeed, numerically  small. 

Now we are ready to study the CP violation convention independent
measure $\overline{\epsilon}$ 
defined in Eq. (\ref{cpp}). Taking into account that in the SM
the phase between $\Gamma_{12}$ and $M_{12}^{LL}$ 
is quite small, ${\cal O}(m_c^2/m_b^2),$ 
we can neglect terms proportional to this phase and obtain 
\bea
 2 \mbox{Re}(\overline{\epsilon}) & \approx & 
\frac{\mid \Gamma_{12}\mid\mid M_{12}^{LR} \mid }{2 \mid 
M_{12}^{LL} \mid^2 \mid 1+\frac{M_{12}^{LR}}{M_{12}^{LL}}\mid^2} 
\sin\sigma ,
\label{elr}
\eea
where
\bea
\sigma = \mbox{Arg}\left(\frac{M_{12}^{LR}}{M_{12}^{LL}}\right).
\label{sigma}
\eea
Note  that this estimate holds for both $B^0_d$ and $B^0_s$ systems
as long as the result is larger than in the SM.
As can be seen, in the left-right model the indirect CP violation
is not related to the small relative phase of  
$\Gamma_{12}$ and $M_{12}^{LL}$ but to the phase between the left
and right contributions to $M_{12}$ which can be appreciable.
Therefore the CP violating 
observables can be significantly enhanced in our model 
when compared with the SM prediction \cite{smeb}
\bea
\mid 2 \, \mbox{Re}(\overline{\epsilon})
\mid \lsim \left\{ \begin{array}{ll}
0.5 \cdot 10^{-3} \;  \; & \;\; B_d^0, \\
0.5 \cdot 10^{-4} \;\; & \;\; B_s^0 .
\end{array}
\right.
\label{esm} 
\eea
To see this enhancement let us assume for a moment that 
$\sin\sigma=1$ and turn later to the analysis of its 
possible values. In this case 
we plot in Fig. 3 the value of $2 \mbox{Re}(\overline{\epsilon}) $ as a
function of $M_2$ fixing the Higgs boson mass to 12 TeV and
in Fig. 4 as a function of $M_H$ fixing the right-handed gauge boson
mass to 1.6 TeV.
With the choosed masses
 left-right symmetry can enhance the CP violation in 
$B_d^0 - \overline{B_d}^0$ system more than a factor of four and 
in $B_s^0 - \overline{B_s}^0$ system almost by two ordes of magnitude.
As seen in the figures, in the left-right symmetric model
$2 \mbox{Re}(\overline{\epsilon}) $ remains larger than the SM prediction 
for quite large values of the right-handed particle masses.
This is a promising result for the future searches of indirect CP violation 
in $B$ physics. 
Since the parameter 
$\overline{\epsilon}$ will be measured from $B$ decay asymmetries, 
an order of magnitude increase in the value of  $\overline{\epsilon}$ 
would mean that hundred times less events are needed
in order to measure it.
 With the designed machine parameters  \cite{bfac} it is not
excluded that the left-right model $\overline{\epsilon}$ 
can  be measured in the planned $B$ factories or hadronic machines,
while the SM $\overline{\epsilon}$, 
taking into account predictions of Eq. (\ref{esm}),
is most probably out of reach in these experiments.  
Therefore, stringent tests of the left-right symmetric model could 
be performed in the near future.

The actual value of the left-right contribution 
(\ref{elr}) will be modified by the phase $\sigma.$ 
Therefore, we have to discuss
how this phase can affect our previous conclusions. 
It follows from \Eq{sigma} that 
\bea
e^{i\sigma} \simeq - \frac{\lambda_t^{RL} \lambda_t^{LR}}
{(\lambda_t^{LL})^2} = - \frac{V_{R,tl} V_{R,tb}^*}
{V_{L,tl} V_{L,tb}^*} \; .
\eea
In the left-right model with spontaneous breakdown of CP,  all the 
CP violating phases in both $V_L$ and $V_R$ depend on a single quantity 
$r \sin\alpha $ ($r=k_2/k_1$). This particular feature makes the 
model very predictive. 
In practice, the phases are calculated analytically only to first 
order in $r \sin\alpha$. However, as  shown in Ref. \cite{frere} 
by numerical calculations, the conventional
small $r \sin\alpha$ approximation gives, indeed,  
very good results. 
An important feature that showed up in these analysis
is the appearance of certain sign factors in the masses. 
According to the studies of the kaon system \cite{seis}, some
combinations of the signs can be ruled out in a phenomenological
basis.

By using the derived formulae for the phases in $V_L$ and $V_R$
to first order in $r\sin\alpha$ one obtains the following approximate 
equations \cite{eg1}
\bea
\sin\sigma^{(d)} &\simeq & \eta_d \eta_b r \sin\alpha \left[
\frac{2 \mu_c}{\mu_s} \left( 1 + \frac{s_1^2 \mu_s}{2 \mu_d} \right)
+ \frac{\mu_t}{\mu_b} \right]
\; ,\nn \\
\sin\sigma^{(s)} &\simeq &\eta_s \eta_b r \sin\alpha \left[ \frac{\mu_c}
{\mu_s} + \frac{\mu_t}{\mu_b} \right] \; ,
\label{signs}
\eea
where
\bea
\mu_i = \eta_i m_i \;\; \;, \; \;\; \eta_i^2=1 \;.
\label{s2}
\eea
It is obvious from Eq. (\ref{signs}) that $\sin\sigma^{(s),(d)}$ depend
strongly on the sign factors in Eq. (\ref{s2}).
However, it is important to notice that both of them
are enhanced by the common dominant factor $m_t/m_b.$ Therefore, 
the values of $\sin\sigma^{(s),(d)}$ can differ maximally by a factor of 2-3
but their order of magnitude is the same. In general, we can write
\bea
\sin\sigma^{(d)} \sim \frac{\eta_s}{\eta_d} 
\sin\sigma^{(s)}.
\eea
At this stage, it is in order to remind some results from the
kaon system analysis. 
While the value of $\eta_d$ is not restricted by the other quark signs,
$\mu_s$ can flip the sign only together with 
$\mu_c$ or $\mu_d$.  
In order to find numerical estimates for $\sin\sigma^{(s),(d)}$
we have to specify the value of $r \sin\alpha.$
It has been shown that, without fine tuning,  the following relation
holds \cite{frere}
\bea
\mid  r \sin\alpha \mid  \lsim \frac{m_b}{m_t}.
\eea
Taking into account the enhancement factor $m_t/m_b$ 
in Eq. (\ref{signs}) one can conclude that $\sin\sigma^{(s),(d)}$
can be naturally of order one as assumed before. On the other hand,
analyses of the kaon system have set lower bounds on 
the value of $r \sin\alpha$ which depend quite strongly on 
the signs of the quark masses. While for  $\eta_s\eta_d = 1$
the lower bound on $r\sin\alpha$ is as low as $5\cdot 10^{-4}$, it is about 
an order of magnitude higher for 
$\eta_s\eta_d =-1$ \cite{seis}.  In the latter case we obtain
\bea
\mid \sin\sigma^{(s),(d)}\mid \gsim 0.2,
\eea
which suggests that the left-right contribution to 
$2\mrm{Re}(\overline{\epsilon})$ is larger than the SM one.
In order to obtain the correct value for the measured $\epsilon_K,$
the values of $r\sin\alpha$ and $M_2$ are correlated.
For small, close to the minimally allowed, values of $r\sin\alpha$
one obtains a very narrow region of $M_2.$ 
For most of the parameter space the values of $\sin\sigma$ in Eq.(\ref{signs})
can be large and
no significant supression of the left-right CP violating
effects  due to the phase $\sigma$ is implied by existing phenomenology.
Unlike stated in previous works, 
enhancement of indirect CP violation in the $B^0_{s,d}$ 
systems in the left-right symmetric models  
can occur in both cases $\eta_s\eta_d =\pm1,$ and particularly for
 $\eta_s\eta_d =-1$. 

\section{Conclusions}

Left-right symmetry provides a promising scenario 
for explaining the origin of
CP violation.  Quite independently of phenomenological
considerations, the left-right symmetric gauge model possesses
the attractive feature that CP can be violated spontaneously
already with the minimal content of the Higgs sector. In this framework,
we have considered the model
where all the CP violating quantities depend on a single phase
$\alpha$ of the scalar bidoublet  vev $\langle \Phi \rangle.$ 
This model has already been 
successfully tested in the kaon system from which
several constraints on the model parameters have been derived.

In the present work we have studied the $B^0-\overline{B}^0$
system in the left-right model with spontaneous breakdown of CP. 
Our  results can be summarized as follows. 
When the right-handed gauge boson and the flavour
changing Higgs boson masses are close to their present lower bounds,
the left-right contribution to the 
$B^0-\overline{B}^0$ mass matrix element $M_{12}$ 
is comparable in magnitude with the SM one.
Assuming that the left-right contribution does not exceed the SM one,
we obtain a new lower bound on the Higgs mass $M_H\gsim 12$ TeV.
However, large cancellations between the two contributions are possible
due to the different phases in the left and right CKM matrices.

While the amount of mixing is 
comparable with  the SM one, CP violating observables 
of the model are much  more promising candidates
to discover  deviations from the SM. The reason is that
the almost common phase of the SM quantities $M_{12}^{LL}$ and 
$\Gamma_{12}$ (the relative phase between these two is very small
suppressing the SM value of $\overline{\epsilon}$) can be quite different
from the phase of the left-right contribution $M_{12}^{LR}$.
Consequently, the convention independent parameter 
$\mbox{{\rm Re}}(\overline{\epsilon}),$ 
which measures the amount of $\Delta B=2$ CP violation,
can be enhanced by a factor of four or more for $B^0_d$ and 
by almost two orders of magnitude for $B^0_s$ systems, 
if compared with the SM prediction.
In terms of 
the requirements on the event multiplicity in the future $B$
factories and hadron facilities 
this would mean that about an order of magnitude less $B_d$ mesons 
should be produced to measure indirect CP violation in the $B$ system. 
Therefore, unlike the SM predictions,
the left-right $\mbox{{\rm Re}}(\overline{\epsilon})$ could
be measured in the future experiments providing a definite signal
of physics beyond the SM.   
According to the  kaon system analyses, no significant 
supression is expected due to the phase $\sin\sigma$ in   
$\overline{\epsilon}.$
An important observation is that CP violating effects for
 $B^0-\overline{B}^0$ can be significant  no matter
whether the left-right and the SM contributions to the
$\epsilon_K$ interfere constructively or destructively.

In conclusion, stringent tests of the left-right symmetric
model can be carried out in the  $B$ meson system when measuring the 
leptonic asymmetries.  

\subsection*{Acknowledgements}

We thank A. Pich and J. Prades for discussions. 
G.B. acknowledges the Spanish Ministry of
Foreign Affairs for a MUTIS fellowship and M.R. thanks the
Spanish Ministry of Education and Culture for a postdoctoral
grant at the University of Valencia. This work is supported by CICYT under 
grant AEN-96-1718.

\newpage

{\large\bf Figure captions}

\vspace{0.5cm}

\begin{itemize}

\item[{\bf Fig. 1.}] 
Feynman diagrams contributing to $|\Delta B|=2$ transition in the 
left-right model.
\\
\item[{\bf Fig. 2.}]
Function $F$ plotted against the right-handed gauge boson mass.
\\
\item[{\bf Fig. 3.}]
$ 2 \mbox{Re}(\overline{\epsilon}) $ as a function of the 
right-handed gauge boson mass for the fixed $M_H=12$ TeV
and $\sin\sigma=1$.
\\
\item[{\bf Fig. 4.}]
$ 2 \mbox{Re}(\overline{\epsilon}) $ as a function of the 
flavour changing Higgs  boson mass for the fixed $M_2=1.6$ TeV
and $\sin\sigma=1$.

\end{itemize}

\newpage

\begin{figure*}[hbtp]
\begin{center}
 \mbox{\epsfxsize=12cm\epsfysize=12cm\epsffile{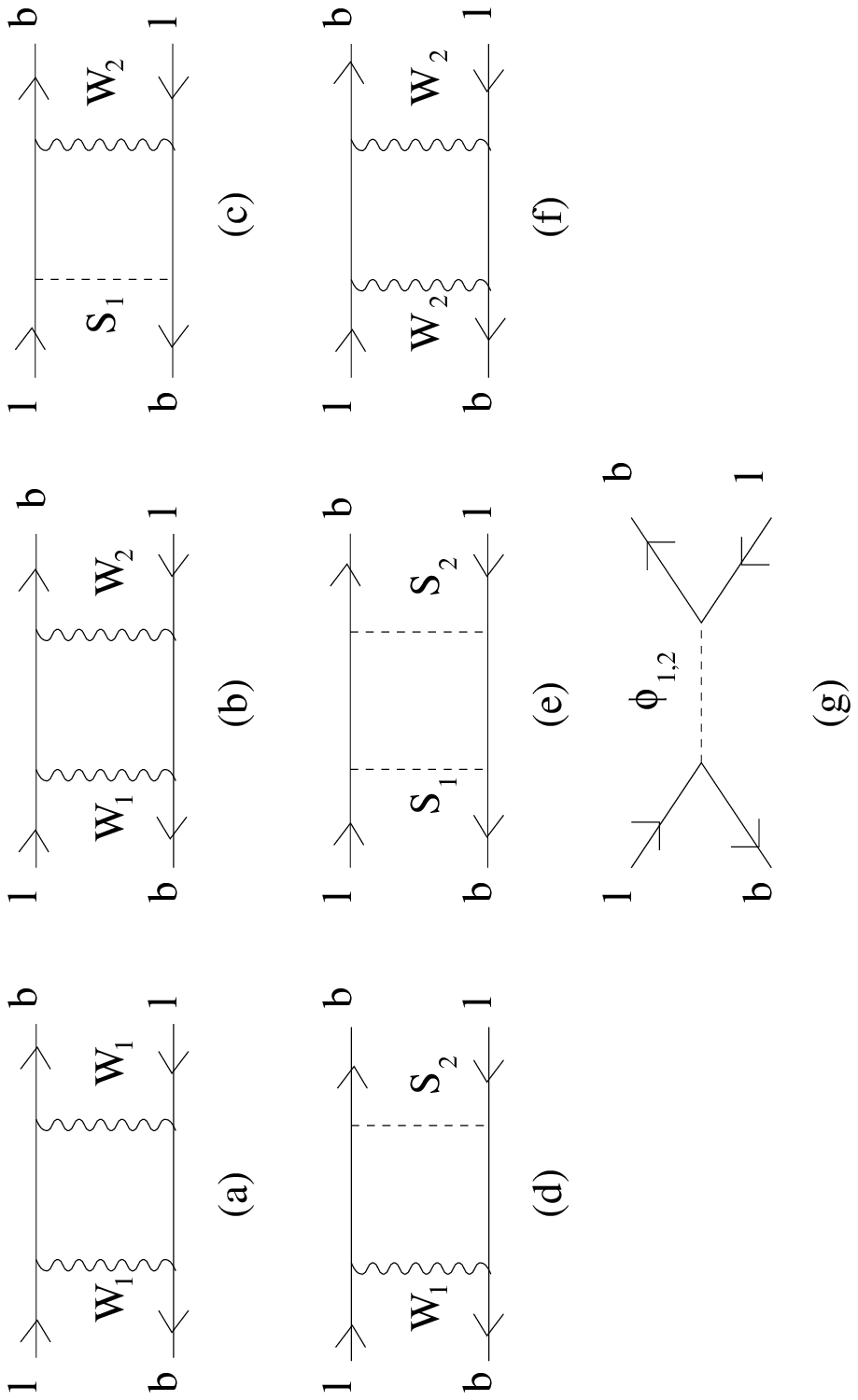}}
\caption{
}
\end{center}
\end{figure*}

\begin{figure*}[hbtp]
\begin{center}
 \mbox{\epsfxsize=12cm\epsfysize=12cm\epsffile{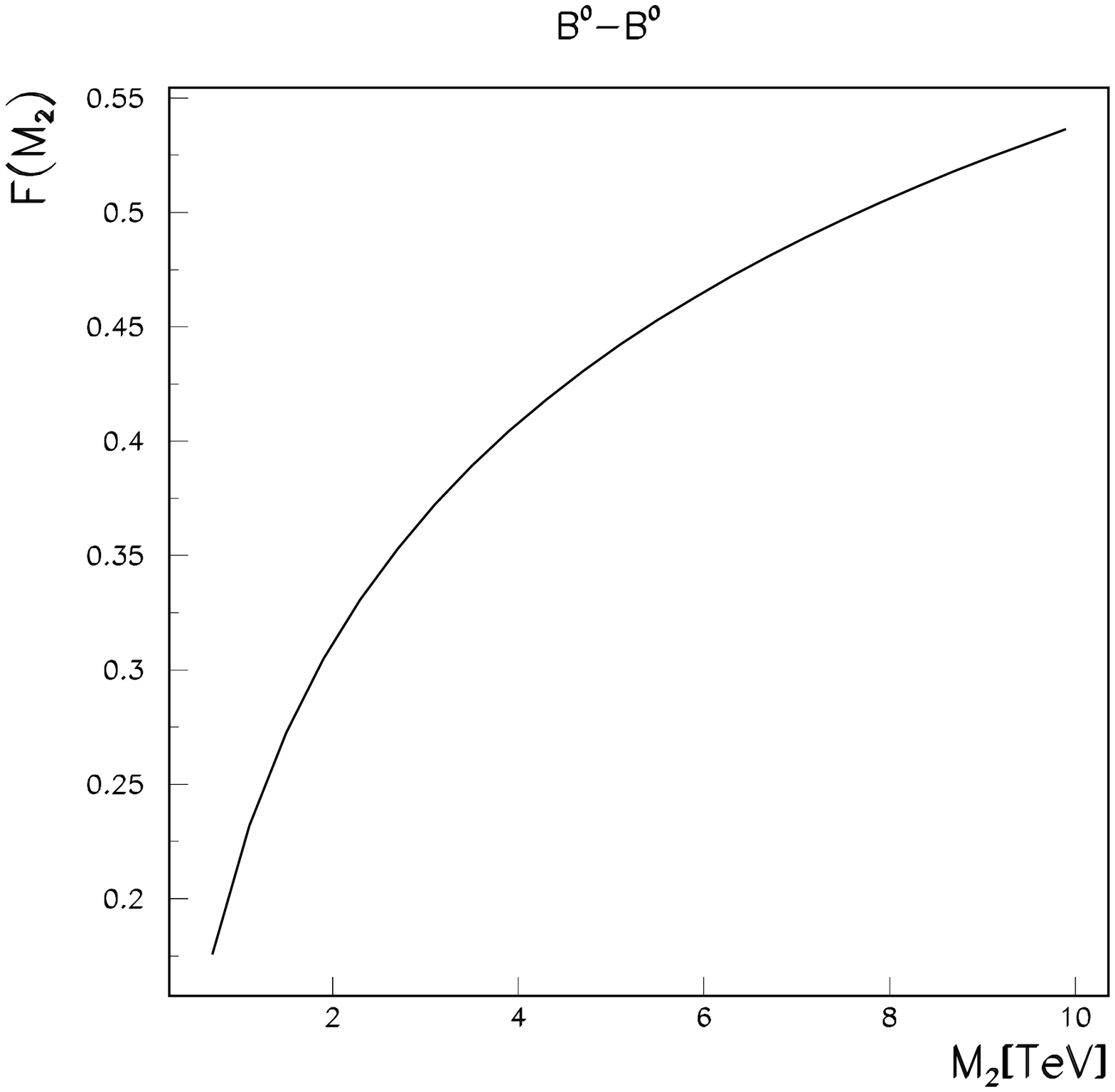}}
\caption{
}
\end{center}
\end{figure*}

\begin{figure*}[hbtp]
\begin{center}
 \mbox{\epsfxsize=12cm\epsfysize=12cm\epsffile{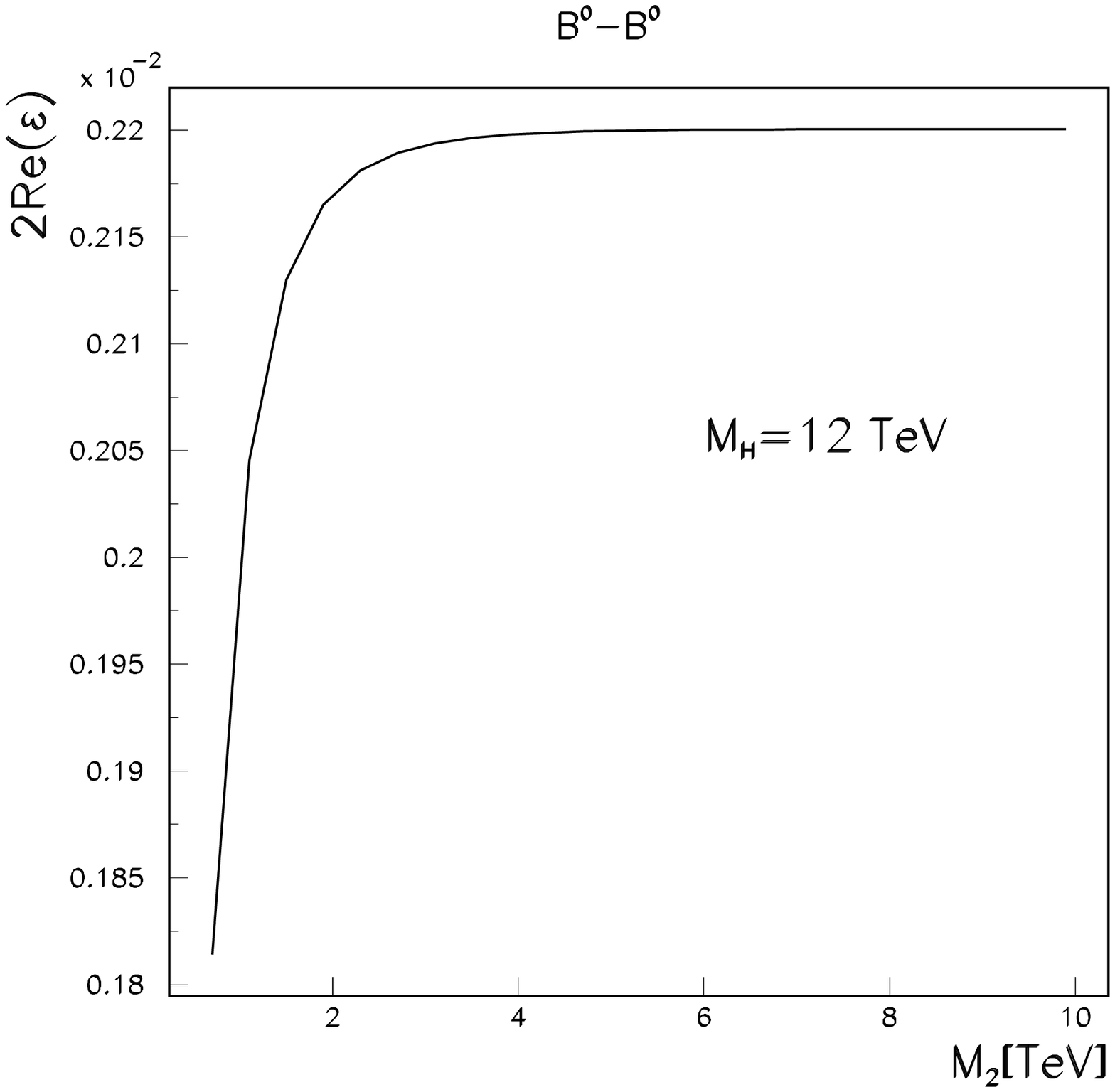}}
\caption{
}
\end{center}
\end{figure*}

\begin{figure*}[hbtp]
\begin{center}
 \mbox{\epsfxsize=12cm\epsfysize=12cm\epsffile{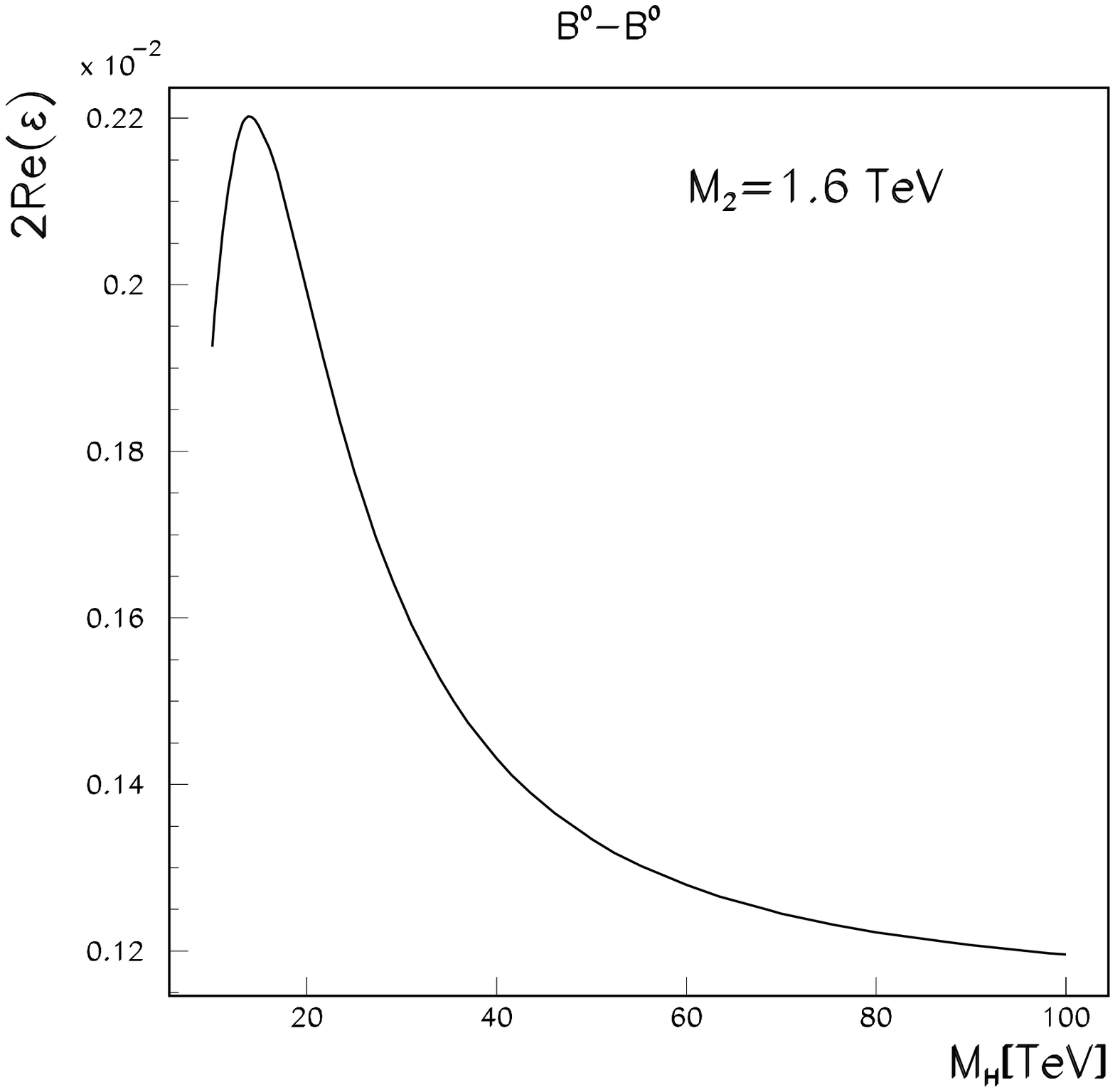}}
\caption{
}
\end{center}
\end{figure*}

\end{document}